\documentclass[aps,prl,superscriptaddress,amsmath,amssymb,twocolumn,nofootinbib]{revtex4}

\usepackage[utf8]{inputenc}
\usepackage{color}
\usepackage{graphicx}
\usepackage{dashbox}
\usepackage{physics}

\def\half{\frac{1}{2}}
\def\kB{k_{\rm B}}
\def\fSMBH{f_{\rm SMBH}}
\def\ODE{\Omega_{\rm DE}}
\def\ODM{\Omega_{\rm DM}}
\def\OM{\Omega_{\rm M}}
\def\Msun{M_\odot}

\begin{document}

\thispagestyle{empty}

\title{Cosmic GREA from SMBH growth}

\author{Juan Garc\'ia-Bellido}\email[]{juan.garciabellido@uam.es}
\affiliation{Instituto de F\'isica Te\'orica UAM-CSIC, Universidad Auton\'oma de Madrid, Cantoblanco 28049 Madrid, Spain}

\date{\today}

\preprint{IFT-UAM/CSIC-23-72}

\begin{abstract}

General Relativistic Entropic Acceleration (GREA) gives a general framework in which to study multiple out-of-equilibrium phenomena in the context of general relativity, like the late accelerated expansion of the universe or the formation of galaxies and the large scale structure of the universe. Here we study the consequences of mass accretion onto massive Black Holes. We find that a population of Super Massive Black Holes (SMBH) whose mass grows significantly due to accretion can act as a source of entropic acceleration and constitute a significant part of the present acceleration of the Universe.

\end{abstract}
\maketitle

\section{I. Introduction}

Understanding the origin of Dark Matter (DM) and Dark Energy (DE) is one of the fundamental quests of Modern Cosmology. Although their phenomenology is well understood, and the actual values of the parameters that characterize these two contributions to the matter and energy content of the universe can be determined to the level of a few percent, their nature is a total mystery.

Primordial Black Holes (PBH) have recently gained a renaissance as a serious contender for all the DM in the universe~\cite{Garcia-Bellido:2017fdg}, when generated during the radiation era from large matter fluctuations that seed small and large scale structures. Those PBH can explain a plethora of astrophysical and cosmological phenomena, and have been suggested to explain the unexpected gravitational wave events seen by LIGO/Virgo~\cite{Garcia-Bellido:2019tvz}. 

Moreover, we have recently proposed that for systems in which there is a production of entropy, the laws of thermodynamics must be incorporated into the Einstein equations via a thermodynamic constraint on the variational principle of the matter-gravity action~\cite{Espinosa-Portales:2021cac}. An effective way of doing this is by adding a viscosity term to the energy-momentum tensor. This new term could give rise to an accelerating universe from the quantum entanglement entropy associated to our cosmological horizon~\cite{Garcia-Bellido:2021idr}, and compared with cosmological observations \cite{Arjona:2021uxs}. However, other sources of entropy could also contribute to the local expansion of the universe.

We pointed out in Ref.~\cite{Espinosa-Portales:2021cac} that in the absence of matter accretion or merging, the conservation of BH entropy cannot account for any cosmic acceleration. However, super massive black holes (SMBH) at the centers of galaxies are surrounded by a massive accretion disk which feeds into the black hole and their masses can grow very fast~\cite{Sijacki:2023kgf}. We consider here the effect that such an early growth in the mass of SMBH has for the entropic force that accelerates the universe. We will conclude that these PBH seeds, which have grown since recombination to become the present SMBH~\cite{Garcia-Bellido:2017fdg}, could also be the source for the Dark Energy of the universe.

\section{II. GREA and the Einstein equations}

The basic concept here is coarsegraining. Suppose we have a mechanical system that consists of two components: i) a set of {\em slow} degrees of freedom described by canonical coordinates with conjugate momenta $(q,\,p)$, and ii) some {\em fast} degrees of freedom coarsegrained as a thermodynamical system characterized by macroscopic quantities, entropy and temperature $(S,\,T)$. The action is then given by
$${\cal S} = \frac{1}{2\kappa}\int d^4x\sqrt{-g}\,R+
\int d^4x\sqrt{-g}\,{\cal L}_m(g_{\mu\nu},\,s)\,,$$
where $s$ is the entropy density and $\kappa = 8\pi G$. The variational principle tells us that
\begin{eqnarray}
    \delta{\cal S} =& {\displaystyle \int d^4x\left(\frac{1}{2\kappa}\frac{\delta(\sqrt{-g}\,R)}{\delta g^{\mu\nu}} +
\frac{\delta(\sqrt{-g}\,{\cal L}_m)}{\delta g^{\mu\nu}}\right)\delta g^{\mu\nu} }\nonumber \\
& + {\displaystyle \int d^4x\sqrt{-g}\, \frac{\partial{\cal L}_m}{\partial s} \delta s\,. }\nonumber
\end{eqnarray}
The interaction between the two components is described by a thermodynamical constraint, in the form of the First Law of Thermodynamics, $\frac{\partial{\cal L}_m}{\partial s} \delta s = \half f_{\mu\nu}\,\delta g^{\mu\nu}$,
which gives rise to the Einstein field equations {\em extended} to out-of-equilibrium phenomena~\cite{Espinosa-Portales:2021cac},
\begin{equation}
	\label{eq:GREA}
	G_{\mu\nu} = R_{\mu\nu} - \half R\,g_{\mu\nu} = \kappa\, (T_{\mu\nu} - f_{\mu\nu}) \equiv \kappa\,{\cal T}_{\mu\nu}\,.
\end{equation}
Here $f_{\mu\nu}$ 
arises from the first law of thermodynamics
\begin{eqnarray}
	\label{eq:TdS}
	-dW = - \vec F\cdot d\vec x &=& dU + \left(p - T\frac{dS}{dV}\right)dV \nonumber \\
 &\equiv& dU + \tilde p\,dV
\end{eqnarray}
where we have defined an {\em effective} pressure $\tilde p$ which reduces to the usual fluid pressure  $p$ in the absence of entropy production. This extra component to the Einstein equations can be interpreted as an effective bulk viscosity term of a real (non-ideal) fluid~\cite{Espinosa-Portales:2021cac}, with $\Theta = D_\lambda u^\lambda$ the trace of the congruence of geodesics,
\begin{equation}
	\label{eq:fmunu}
	f_{\mu\nu} = \zeta\,\Theta \,(g_{\mu\nu}+u_\mu u_\nu) = \zeta\,\Theta\,h_{\mu\nu} \,,
\end{equation}
such that the covariantly-conserved energy-momentum tensor has the form of a perfect fluid tensor,
\begin{eqnarray}
	\label{eq:Tmunu}
	{\cal T}^{\mu\nu} &=& p\,g^{\mu\nu} + (\rho + p)u^\mu u^\nu -  \zeta\,\Theta\,h^{\mu\nu} \\ &=& \tilde p\,g^{\mu\nu} + (\rho + \tilde p)u^\mu u^\nu\,,
\end{eqnarray}
and, imposing the thermodynamic constraint~(\ref{eq:TdS}), the bulk viscosity coefficient $\zeta$ can be written as
\begin{equation}
	\label{eq:zeta}
	\zeta = \frac{T}{\Theta}\frac{dS}{dV}\,.
\end{equation}
In the case of an expanding universe, $\Theta=\frac{d}{dt}\ln V = 3H$ and the coefficient becomes $\zeta = T\dot S/(9H^2a^3)$, see~\cite{Garcia-Bellido:2021idr}, with $S$ the entropy per comoving volume of the Universe. Entropy production therefore implies $\zeta > 0$.

Note that the energy-momentum tensor is still diagonal, ${\cal T}^\mu_{\ \ \nu} = {\rm diag}(-\rho,\,\tilde p,\,\tilde p,\,\tilde p)$, and that the $00$ component is unchanged with respect to GR. Only the $ij$ com\-ponent has the entropy-growth dependence via $\tilde p$.

The Raychaudhuri equation~\cite{Wald:1984rg} for geodesic motion ($a^\mu = u^\nu D_\nu u^\mu = 0$) in the absence of shear ($\sigma_{\mu\nu}\,\sigma^{\mu\nu}=0$) and vorticity ($\omega_{\mu\nu}\,\omega^{\mu\nu}=0$) is given by
\begin{eqnarray}
	\label{eq:Ray} \nonumber
	\frac{D}{d\tau}\Theta + \frac{1}{3}\Theta^2 &=&\!- R_{\mu\nu} u^\mu u^\nu \\ &=&\!- \kappa\left(T_{\mu\nu}u^\mu u^\nu + \half T^\lambda_{\ \,\lambda} - \frac{3}{2}\zeta\Theta\right) \\ &=&\!- \frac{\kappa}{2}(\rho + 3\tilde p) = -\frac{\kappa}{2}\left(\rho + 3p - 3T\frac{dS}{dV}\right) \nonumber\,.
\end{eqnarray}
Due to the extra entropic term in the effective pressure $\tilde p$, even for matter that satisfies the strong energy condition, $\rho + 3p > 0$, it is possible to prevent gravitational collapse, i.e. $\dot\Theta + \Theta^2/3 > 0$, as long as the production of entropy is significant enough, $3TdS/dV > (\rho + 3p) > 0$.

\section{III. Entropy of the BH horizon}

One can also wonder about the effect of the entropy associated to space-time itself, in particular to horizons. It can be incorporated in a natural way by extending the Einstein-Hilbert action with a surface term, the Gibbons-Hawking-York (GHY) term of Refs.~\cite{York:1972sj,Gibbons:1976ue}.

Let us consider a space-time manifold $\mathcal{M}$ with metric $g_{\mu \nu}$, which has a horizon hypersurface that we denote by $\mathcal{H}$. This is a submanifold of the whole space-time. By taking  $n^{\mu}$, the normal vector to the hypersurface $\mathcal{H}$, we can define an inherited metric on $\mathcal{H}$:
\begin{equation}
    h_{\mu \nu} = g_{\mu \nu} + n_{\mu} n_{\nu}\,.
\end{equation}
With this, one can define the GHY term as
\begin{equation}
    S_{\rm GHY} = \frac{1}{8\pi G} \int_{\mathcal{H}} d^3 y \sqrt{h} K\,,
\end{equation}
where $K$ is the trace of the extrinsic curvature of the surface. Notice that we are not foliating the entire space-time, but rather considering the properties of a particular hypersurface, the horizon.
From the thermodynamic point of view, the GHY term contributes to the free energy of the system. Hence, it can be related to the temperature and entropy of the horizon as~\cite{Espinosa-Portales:2021cac}
\begin{equation}
    S_{\rm GHY} = - \int dt \,N(t)\,T S\,.
\end{equation}
where we have kept the lapse function $N(t)$, to indicate that the variation of the total action with respect to it will generate a Hamiltonian constraint with an entropy term together with the ordinary matter-energy terms. This leads to the realization that what drives gravity in a thermodynamical context is not just the internal energy of a system, $U$, but its Helmholtz free energy, $F=U-TS$. In other words, entropy gravitates, or perhaps we should better say entropy ``antigravitates" since it is responsible for a repulsive force. What gravitates is information.

\subsubsection{Schwarzschild black hole}

In order to illustrate this, let us now compute the GHY action for the event horizon of a Schwarzschild black hole of mass $M$. Its space-time is described by the metric:
\begin{equation}
    ds^2 = - \left[1- \frac{2GM}{r} \right] \!dt^2 + \left[1- \frac{2GM}{r} \right]^{-1}\!\!dr^2 + r^2 d\Omega^2\,.
\end{equation}
The normal vector to a 2-sphere of radius $r$ around the origin of coordinates is
\begin{equation}
    n = - \sqrt{1 - \frac{2GM}{r}} \partial_r\,.
\end{equation}
With this, the trace of the extrinsic curvature for such a sphere scaled by the metric determinant is
\begin{equation}
    \sqrt{h} K = (3GM - 2r) \sin\theta\,.
\end{equation}
Integrating over the angular coordinates and setting the 2-sphere at the event horizon, i.e. $r = 2GM$, and restoring for a moment $c$, the GHY boundary term becomes
\begin{equation}\label{eq:GHYaction}
    S_{\rm GHY} = - \frac{1}{2} \int dt \,Mc^2  = - \int dt \,T_{\rm BH} S_{\rm BH}\,,
\end{equation}
where $T_{\rm BH}$ is the Hawking temperature and $S_{\rm BH}$ is the Bekenstein entropy of the Schwarzschild black hole~\cite{Hawking:1974sw},
\begin{equation}\label{eq:Hawking}
    \kB T_{\rm BH} = \frac{\hbar c^3}{8\pi G M} \,, \hspace{9mm} S_{\rm BH} = \kB\,\frac{4\pi G M^2}{\hbar c}\,.
\end{equation}
This favors the interpretation of the GHY boundary term as a contribution to the Helmholtz free energy in the thermodynamic sense, $F=U-TS$. 

Note also that the action (\ref{eq:GHYaction}) is classical, essentially the rest mass energy of the BH, although both the Hawking temperature and Bekenstein entropy are quantum mechanical quantities, associated with the entanglement of the fundamental degrees of freedom between the interior and exterior of the horizon of a black hole. We interpret this result as being an {\em emergent} phenomenon, from microscopic degrees of freedom to a coarsegrained description in terms of thermodynamical quantities like temperature and entropy, where all fundamental constants ($\hbar, \kB, G$) cancel out, except $c$.

In Ref.~\cite{Espinosa-Portales:2021cac} we argued that, in the absence of significant clustering or merging, the masses of black holes remained constant and thus there would be no entropy production associated with stellar black holes in our universe. However, let us consider here an alternative possibility.

\section{IV. PBH at the origin of both \hspace{2cm} Dark Matter and Dark Energy}

We consider here the possibility that Dark Matter is composed of Primordial Black Holes (PBH) and that a small fraction of these black holes, with masses $M_{\rm BH} \sim10^6\Msun$, constitute the seeds of SMBH at the centers of galaxies~\cite{Garcia-Bellido:2017fdg}. These black holes accrete mass from the environment at a rate that is conmensurate with the rate of expansion of the universe. When accretion of gas from the surroundings reaches the Eddington limit, the mass of the SMBH grows like~\cite{Carr:2023tpt}
\begin{equation}\label{eq:Eddington}
    \dot M = \frac{4\pi G\,m_p}{0.1 c\,\sigma_T}\,M \simeq 
    \frac{M}{40\,{\rm Myr}} = \frac{2}{t(z_*)}\,M\,,
\end{equation}
where $m_p$ is the proton mass and $\sigma_T$ is the Thomson cross-section, and we have used $t(z_*\simeq35) = 80$ Mpc for a Universe with $\OM=0.31$. If we now assume that SMBH continue to accrete gas at the Eddington limit with a rate that decreases in time with the density of matter available in the universe ($\dot\rho/\rho = - 2/t$), then the mass of SMBH will grow due to accretion as $M\propto t^2 \sim a^3 = V$ in the past, at least since $z_*\simeq35$, see also~\cite{Farrah:2023opk}. If we compute the general relativistic entropic acceleration (GREA) induced by the growth of entropy associated with this increase in mass,  $V\, dS_{\rm SMBH} = 2\,S_{\rm SMBH} dV$, we see that it contributes with a {\em constant} negative pressure
\begin{equation}\label{eq:Growth}
    p_S = -T\frac{dS}{dV} = - 2\frac{TS}{V} = - \frac{N_{\rm SMBH}M_{\rm SMBH}}{V} = - \rho_{\rm SMBH}\,,
\end{equation}
where the total entropy is 
$$S = \sum_i S^{(i)}_{\rm SMBH} = N_{\rm SMBH}S_{\rm SMBH},$$ 
with $N_{\rm SMBH}$ the total number of SMBH in the universe, assumed {\em constant} (i.e. without SMBH merging). We then find the Raychaudhuri equation~(\ref{eq:Ray}), where we have separated the ordinary (adiabatic) matter characterized by $(\rho,\,p)$ from the SMBH,
\begin{eqnarray}
	\label{eq:GREA} \nonumber
	\dot H + H^2 = \,\frac{\ddot a}{a}\!&=&\!- \frac{4\pi G}{3}\left( \rho + 3p + \rho_{\rm SMBH} + 3p_S\right)\\ 
    &=&\!-\frac{4\pi G}{3}\left( \rho + 3p\right)
    + \frac{8\pi G}{3}\rho_{\rm SMBH}\,.
\end{eqnarray}
The last {\em constant} term can be interpreted as an {\em effective} and {\em positive} Cosmological Constant term $\Lambda=\kappa\,\rho_{\rm SMBH}$, 
driving cosmic acceleration, while the rest of the (adiabatic) matter and radiation in the universe, the baryons and photons, as well as the PBH that do not accrete significantly and thus act as Cold Dark Matter, would contribute to cosmic deceleration.

We can now evaluate the contribution of such a term to the present acceleration of the universe. If PBH constitute all of the Dark Matter in the universe, and a small fraction of these (the seeds of SMBH) have masses that increase with the cosmic volume, then their contribution is identical to that of a cosmological constant with the same density,
\begin{equation}\label{eq:HamiltonianConstraint}
    H^2 = \frac{8\pi G}{3}\left(\rho + \rho_{\rm PBH} + \rho_{\rm SMBH}\right)\,.
\end{equation}
What these equations (\ref{eq:GREA}) and (\ref{eq:HamiltonianConstraint}) are telling us is that primordial SMBH, rather than contributing to the Dark Matter of the universe today, they are actually the source of Dark Energy. Whether there has been a gradual change from DM to DE over the course of time, as PBH grow due to accretion, is a matter of discussion when compared with cosmological observations~\cite{Arjona:2021uxs}.

There is also the possibility that only a small fraction of the PBH that contributed to DM in the early universe, e.g. just the SMBH in the centers of galaxies, grew sufficiently rapidly to contribute  significantly to the GREA of the universe. In that case, the bulk of the PBH would still contribute to DM today and only a small fraction, around $\fSMBH = 5\times10^{-5}$ of all PBH~\cite{Cappelluti:2021usg}, would contribute to DE in the form of rapidly accreting SMBH with entropy growth associated with their horizons (\ref{eq:Growth}).

The rapid increase in mass of the SMBH at the centers of galaxies since $z_*\simeq 35$ can quickly increase their contribution to DE (and compensate for their tiny contribution in numbers to the total amount of PBH),
\begin{equation}\label{eq:DEfromDM}
    \ODE = \fSMBH \,\ODM \,(1+z_*)^3 = 0.69\,,
\end{equation}
for $\ODM=0.26$. A more sophisticated computation may be needed for the case that PBH in a broader range of masses happen to accrete mass at a slower rate but for a longer time,
\begin{equation}\label{eq:DEfromDM}
    \ODE = \ODM \int \frac{f(M)}{M}\,\frac{dM}{dz} dz\,,
\end{equation}
where $f(M)$ is the fraction of DM in the form of PBH that accrete mass with a rate $dM/dz$. The integral of all contributions would have to be used to estimate not only the actual value of $\ODE$, but also its possible rate of change, and thus the effective DE parameters $(w_0,\, w_a)$, which could then be compared with observations~\cite{Arjona:2021uxs}. Such a computation is beyond the scope of this letter.

Note that we have assumed above that the SMBH are uniformly distributed in our universe when they started accreting gas from their surroundings at $z_*\sim35$. Their density at that time, $\rho_{\rm SMBH} \simeq \rho_c^{0}\,\fSMBH\,\ODM \,(1+z_*)^3 = 10^6\,M_\odot /(20\,{\rm kpc})^3$, is about a SMBH within a sphere of 20 kpc radius, which would correspond to a comoving radius of less than a Mpc today, well below the scale of inhomogeneities. Therefore, we can ignore possible inhomogeneities in the local GRE acceleration induced by the distribution of SMBH and their mass growth. 

\section{V. Conclusions}

We have explored in this letter the possibility that GREA may account for the present acceleration of the universe arising from the entropy growth associated with the mass accretion onto SMBH since the cosmic dark ages. This way, a tiny fraction of the PBH that seed structure and contribute to the Dark Matter of the universe, more specifically the SMBH at the centers of galaxies seeded by the massive PBH from the $e^+e^-$ annihilation era~\cite{Carr:2019kxo}, would gain mass and drive also the acceleration of the universe.

It is only recently that SMBH, seeded by massive PBH, have started to accrete mass and induce a cosmic acceleration via the entropic force associated with GREA. In the past, only matter and radiation drove the decelerated expansion of the universe. When GREA started to drive acceleration via SMBH growth, it acted as an effective cosmological constant term, which eventually dominated the free energy budget. This could explain the actual value of the so-called dark energy density today, as well as the coincidence problem, i.e. why both dark energy and dark matter contributions are of the same order.

The local GREA around each SMBH is sufficiently uniformly distributed in the past (when mass growth from accretion was dominant), over comoving scales of order a Mpc, that we can assume homogeneity of the accelerated expansion over the entire universe. 
Moreover, GREA from the cosmological causal horizon~\cite{Garcia-Bellido:2021idr} is still a valid alternative. Both come from emergent phenomena associated with horizon entropies; they both have a quantum origin in entanglement, and they could have comparable contributions to the present acceleration of the universe. Which one dominates today and how does this splitting determine the rate of change of acceleration over time, and thus observations of the effective Dark Energy parameters $(w_0,\, w_a)$, is still a matter of investigation.

What is the fate of the SMBH's contribution to DE? There will be a time in which mass growth of SMBH will stop, after consuming the majority of the gas in their accretion disks. From then onwards, SMBH will conserve entropy (unless they merge with other SMBH), and thus the associated GREA will stop being a driving accelerating force. In fact, this epoch of stalled mass growth may be near the present age of the Universe. These quiescent supermassive black holes will then only contribute as CDM, like the rest of PBH, and thus will decelerate the expansion of the universe, which will redshift away their energy density and will end in an empty Universe (possibly after the evaporation of all these BH due to Hawking radiation~\cite{Hawking:1974sw}), corresponding to a Minkowsky space-time.


\section{Acknowledgements}

The author acknowledges support from the Spanish Research Project PID2021-123012NB-C43 [MICINN-FEDER], and the Centro de Excelencia Severo Ochoa Program CEX2020-001007-S at IFT. 

\bibliography{main}

\end{document}